\documentclass[a4paper,fleqn]{cas-dc}

\usepackage[numbers]{natbib}
\usepackage{orcidlink}
\usepackage{amsmath}
\usepackage{graphicx} 

\def\tsc#1{\csdef{#1}{\textsc{\lowercase{#1}}\xspace}}
\tsc{WGM}
\tsc{QE}

\begin{document}
\let\WriteBookmarks\relax
\def\floatpagepagefraction{1}
\def\textpagefraction{.001}

\shorttitle{Optimal price signal generation}    

\shortauthors{S. S. Tohidi et~al.}  

\title[mode = title]{Optimal price signal generation for demand-side energy management}  

\makeatletter\def\Hy@Warning#1{}\makeatother



%

\author[1]{Seyed Shahabaldin Tohidi,\orcidlink{0000-0002-4566-667X}}[orcid = 0000-0002-4566-667X]

\cormark[1]


\ead{sshto@dtu.dk}


\affiliation[1]{organization={Department of Applied Mathematics and Computer Science, Technical University of Denmark},
            city={Kgs. Lyngby},
            postcode={DK-2800}, 
            country={Denmark}}

\author[1]{Henrik Madsen}[orcid = 0000-0003-0690-3713]


\ead{hmad@dtu.dk}

\author[1]{Davide Calì}


\ead{dcal@dtu.dk}



\author[1]{Tobias K. S. Ritschel}[orcid = 0000-0002-5843-240X]


\ead{tobk@dtu.dk}





\begin{abstract}
    Renewable Energy Sources play a key role in smart energy systems. To achieve 100$\%$ renewable energy, utilizing the flexibility potential on the demand side becomes the cost-efficient option to balance the grid. However, it is not trivial to exploit these available capacities and flexibility options profitably. The amount of available flexibility is a complex and time-varying function of the price signal and weather forecasts. In this work, we use a Flexibility Function to represent the relationship between the price signal and the demand and investigate optimization problems for the price signal computation. Consequently, this study considers the higher and lower levels in the hierarchy from the markets to appliances, households, and districts. This paper investigates optimal price generation via the Flexibility Function and studies its employment in controller design for demand-side management, its capability to provide ancillary services for balancing throughout the Smart Energy Operating System, and its effect on the physical level performance. Sequential and simultaneous approaches for computing the price signal, along with various cost functions are analyzed and compared. Simulation results demonstrate the generated price/penalty signal and its employment in a model predictive controller.
\end{abstract}


\maketitle

\section{Introduction}\label{sec:intro}
Adoption of Renewable Energy Sources (RESs) is a prominent step toward carbon neutralization, and consequently, is a solution to mitigate global warming. Connecting these new sources to the grid can bring new challenges due to their intermittence, fluctuating power generation, and dependency on 
environmental conditions \cite{Lund15, Li22}. For instance, the output of wind turbines and photovoltaic (PV) panels is a complex function of many variables, like meteorological variables and dirtiness of blades or panels, and varies seasonally, daily, or at even higher frequencies. Difficulty in predicting these sources of power generation complicates the energy management \cite{Rasmussen20, Sorensen23a, Sorensen23b}.

Different from the traditional energy management systems, which are dependent on increasing the electricity supply to overcome the demand peaks, modern energy systems rely on demand-side management (DSM). To this end, demand loads are accommodated to the supply capacity throughout each hour of the day. This requires a permanent data transfer between the supply and demand sides \cite{Pean19, Pean17, pandey2021hierarchical, tsaousoglou2021demand, Toquica21, Murthy22} as well as a hierarchy of methods and models for aggregation, forecasting, control and optimization \cite{madsen2015a}. The entire hierarchical framework, which is entitled the Smart Energy OS (Operating System) (\cite{madsen2016a, parvizi2020a}), has been used in several projects to activate demand-side flexibility; see for instance \cite{madina2019a, dognini2022a,blomgren2022a}. 

\begin{figure*}
\centering
\includegraphics[width=0.55\textwidth]{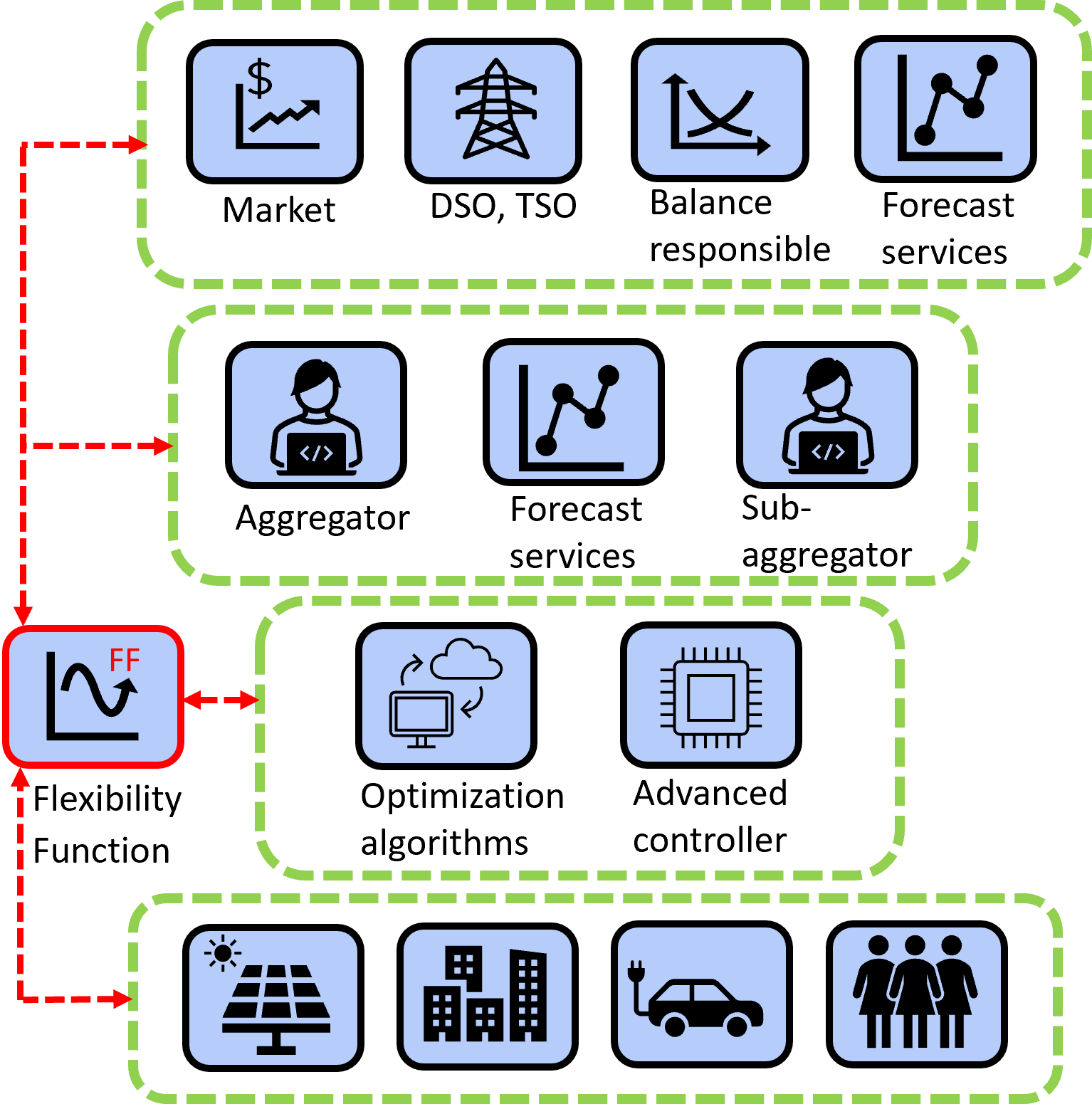}
\caption{The Smart Energy OS.}
\label{Fig:SEOS}
\end{figure*}

Various levels and elements of Smart Energy OS are depicted in Figure \ref{Fig:SEOS}. The required data and information transfer between the elements of this framework leads to a high level of complexity and indicates the need for a digitalization of the entire energy domain as offered by the Smart Energy OS. For the modern weather-driven energy system, methods for connecting high-level grids and balancing challenges with the low-level flexible demand are needed \cite{Nys21}. This can be offered efficiently by enhancing the connection between the elements using a reliable model for the price-demand relationship 
\cite{corradi2013a,Alpagut22, Guillen22}. 

A Flexibility Function (FF) is introduced as a key element to keep the different parts of Smart Energy OS connected. The FF is a stochastic model that represents the price-demand dynamics in energy systems. On one hand, it provides information on the load prediction and flexibility potential for aggregators, grid operators, and balance-responsible parties, and on the other hand, it is capable of generating price signals for the electricity market and advanced controllers in energy management systems \cite{Rune18, Rune20, tsaousoglou2022market, Domink20}. In this paper we will focus on power systems, but the Smart Energy OS is able efficiently to handle integrated energy systems and sector coupling. It is rather obvious that sector coupling like power to heat and PtX, enhances the flexibility of the energy system and hence the possibilities for large-scale integration of fluctuating renewables.  

It is noted that when employing a nonlinear price-demand dynamical system, like the Flexibility Function, in various levels of the Smart Energy OS a methodology to guarantee stability is required. 
To take care of dynamic updates of FF, an Adaptive FF (AFF) is proposed in \cite{Tohidi2023adaptive}. The AFF takes the time-variation of price-demand dynamics into consideration and updates the price signal such that the error between the actual and predicted demand is minimized while stability is guaranteed.

In this paper, we investigate
optimal price generation via the FF and study its employment in controller design
for demand-side management. The generated price/penalty signal can then be utilized for ancillary
services throughout the Smart Energy OS. In particular, we focus on the Smart Energy OS and its elements and their connections and introduce the capabilities of FF to enhance it. For instance, we discuss demand predictability as a function of price, FF for grid balancing, and FF for physical level performance improvement. Sequential and simultaneous approaches for computing the price signal and providing ancillary services, along with various cost functions are analyzed and compared. The generated optimal price is then utilized in a model predictive controller to improve cost efficiency by shifting the electricity load to some low electricity price periods of the day. 

This paper is organized as follows: Section~\ref{sec:SEOP} gives an overview of a Smart Energy OS and its components. Section~\ref{sec:linking} describes the connection between various elements of the energy system. In Section~\ref{sec:DemandPrediction}, demand prediction using FF is introduced. In Section~\ref{sec:OptPrice}, optimal price generation using FF is investigated. Sections \ref{sec:HighLevel} and \ref{sec:LowLevel} discuss the benefits of using FF for the higher and lower levels of energy systems, respectively. Section~\ref{sec:Simul} demonstrates the simulation results of optimal price generation. In~Section~\ref{sec:Out}, we discuss how price signal generation based on FFs can be used to indirectly control other types of energy demand, and we use district heating as an example. Finally, a summary is given in Section~\ref{sec:Sum}.









\section{Smart energy operating system}\label{sec:SEOP}
Smart energy systems and their elements for forecasting, control, and optimization constitute the so-called Smart Energy OS (SE-OS). The SE-OS describes the connection and data transfer between each segment of the power system, decision-making mechanisms, stability algorithms, etc.

Various elements and levels of SE-OS are given in Figure \ref{Fig:SEOS}. This figure divides the elements of the Smart Energy OS into four levels. The top level consists of electricity markets,  Transmission System Operators (TSOs), Distributed System Operators (DSOs), and balance responsible parties (BRPs). Generally, energy management at this level is done for a country, region, or city. Their macroscopic point of view takes care of stability, performance, and electricity prices. 

Aggregators and related forecasting services are in the next level of OS. They are responsible for the energy management in a city, district, micro-grid, or neighborhood and provide vital information required for energy management and control at this medium level. Neighborhoods and buildings, renewable energy resources like large-scale PVs, and wind turbines can also be considered at this level. Humans, appliances, batteries, and charging panels for electric vehicles (EVs) are at the lower level. Heating, ventilation, and air conditioning (HVAC) systems and storage tanks are also parts of the lower level.  

Within the Smart Energy OS framework advanced controllers and optimization algorithms are employed to preserve the stability and optimize the performance of the lower level. As a result of the market clearing, optimization, and controllers at the upper and medium levels, a dynamic price signal is provided for the low-level controllers. This price signal is a composite signal resulting from all the market, balancing, and ancillary service problems at the higher level. The objective of the lower-level controllers is to activate demand-response solutions, such that the efficiency of the entire system is guaranteed. 

The concept offered by the Smart-Energy OS provides a real-time one-way broadcasting of dynamic prices. The information embedded in the FF can be harvested either offline or by edge computing, e.g. in the smart-home management system. This provides solutions with a focus on cyber security and the one-way real-time signal ensures privacy by design. 

As seen by the end-users, i.e. both the industry and the residential sector, the setup ensures that the end-users are in charge of making the final decisions, and the framework aims at facilitating a trusted spatial-temporal setup that puts priorities in empowering the users such that they are able to provide digitalized and efficient demand-response solutions without being subject to disproportionate technical requirements, contracts, administrative requirements, charges, and procedures.

\section{Linking market's level to the physical level}\label{sec:linking}
One objective of the Smart Energy OS is to facilitate load shifting through the utilization of flexible assets, e.g., buildings, supermarkets, and water treatment plants. The operators of such flexible assets can offer their flexibility through conventional market mechanisms. However, load shifting only has a low priority for many flexible assets. Therefore, they estimate their flexibility conservatively. For instance, although a supermarket is able to shift its cooling load from peak hours, its main priority is to prevent its goods from being spoiled. Consequently, in the presence of uncertain weather forecasts, they are unlikely to bid into markets and fully exploit their potential flexibility. Similarly, wastewater treatment plants can be operated flexibly, but the primary objective is to prevent overflow and the secondary objective is to sustain the active part of the sludge by limiting the flow rate. Therefore, even for low probabilities of severe rainfall, they are also unlikely to bid into markets. 

For the lower levels of the hierarchy, we consider conventional market mechanisms to be unfeasible. For instance, the low-voltage grid operator (DSO) has to ensure a proper voltage level throughout the electrical feeder and ensure that the temperature of the transformers is within the given constraints. However, typically the number of potential market participants along a feeder is too low to ensure enough bids for a conventional market, and furthermore, it is often seen as a challenge e.g. for residential users to provide bids or sign a flexibility contract. Here the simple setting of a dynamic broadcasting of prices is often successful in activating the flexibility potential \cite{madina2019a}.


Operationally the utilization of flexibility can be mitigated by letting specialized aggregators trade on the electricity market and broadcast price signals to the flexible asset operators. As the aggregators offer the same price to all of their customers, this approach is fair and transparent. Specialized aggregators could focus on specific sectors. As an example, it seems to be advantageous to have specialized aggregators for harvesting the flexibility of wastewater treatment plants, since this calls for specific knowledge of the physical systems delivering the flexibility.  
However, it requires that each aggregator can assess and predict the flexibility of the underlying assets. The FF is used for this purpose. It quantifies the dynamic load behavior as a function of price signals, i.e., the temporal price-sensitivity of the flexible asset. Furthermore, 1)~it is data-driven and should continuously be updated based on the latest data from the asset, and 2)~it can account for the uncertainty in the operator's behavior. We refer to the work by \cite{Rune18, Rune20} for more details and different types of FFs.
In summary, FFs constitute a central element of the Smart Energy OS as it provides the link between markets (where the aggregator trades) and the physical reality (where the flexible asset is operated).




\section{Demand prediction using FF}\label{sec:DemandPrediction}

The ability to predict demand is an appealing and important task in energy systems. Having enough information about the demand leads to more efficient energy management. For instance, this information can be employed in advanced controllers for peak shaving and load-shifting purposes. Consequently, this invaluable information is effective in diminishing the costs of energy consumption dramatically. 

The FF, as a dynamic mapping between demand and price, is capable of providing demand prediction as a function of baseline demand and the electricity price. The nonlinear FF, as introduced by \cite{Rune20}, is a stochastic dynamical system of the form
\begin{align}\label{eq:1}
    d{X}_t &= f(X_t, U_t, B_t)dt + g(X_t)dw,\notag \\ 
    Y_t &= h(X_t, U_t, B_t)
\end{align}
where $X_t$ is the flexibility state, $U_t$ is the electricity price, $B_t$ is the baseline energy demand, and $w$ is a Wiener process. The output of the nonlinear FF is the predicted demand, $Y_t$.

\begin{figure}
\centering
\includegraphics[width=0.5\textwidth]{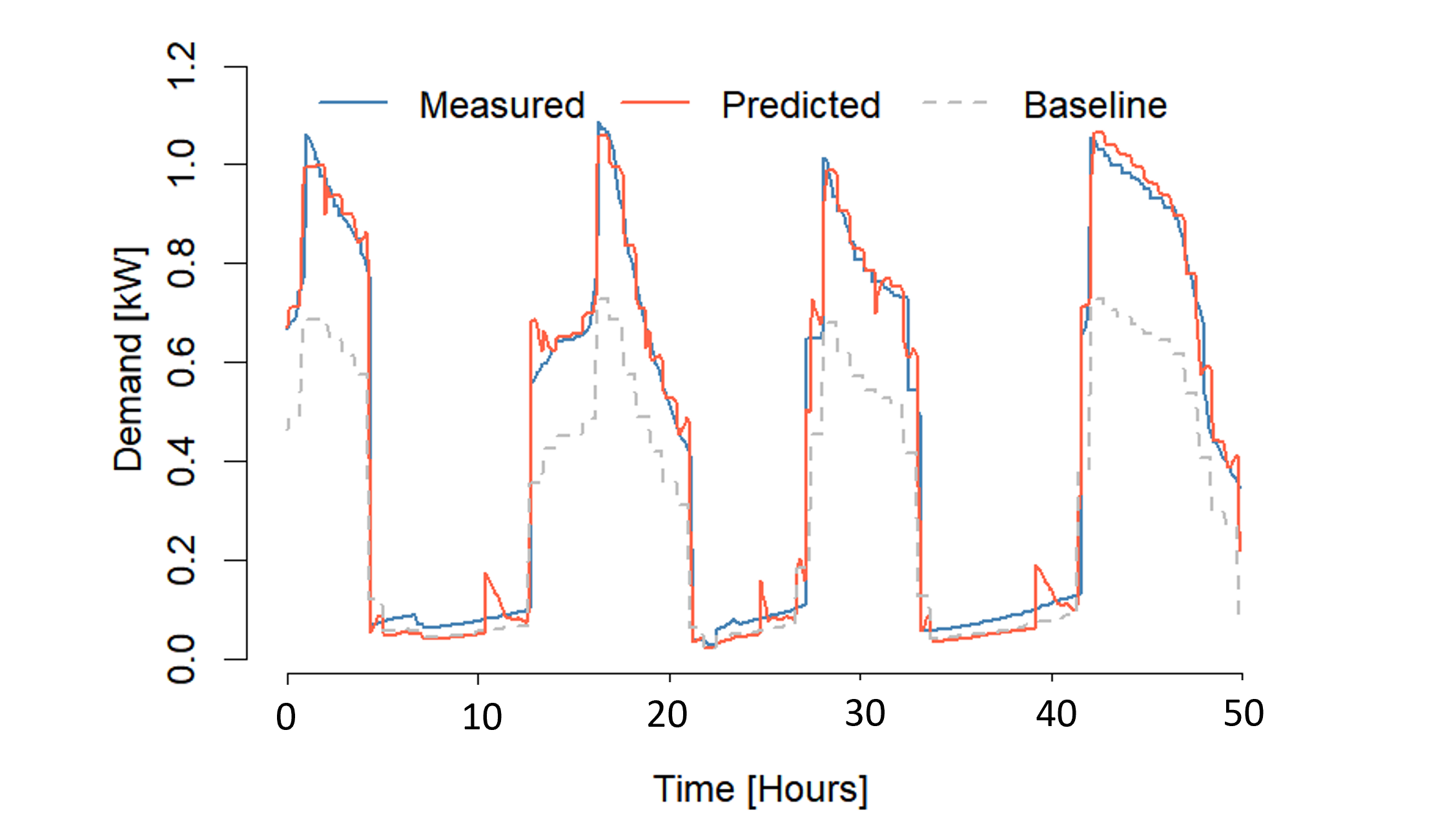}
\caption{Demand prediction using flexibility function}
\label{Fig:FFpredict}
\end{figure}

Figure \ref{Fig:FFpredict} demonstrates the predictability of a nonlinear FF provided for a new development in Fredrikstad, Norway \cite{Salom21}. The demand predicted by the FF is close to the measured demand. This makes the FF a dynamic demand prediction tool that can be employed at different levels of the Smart Energy OS. Notice, that the FF provides a method for predicting the demand for price-responsive systems in the case of dynamic pricing mechanisms. 

The ability to predict the demand based on the electricity price makes the FF a key element for generating optimal price (penalty) signals, that can be utilized in many control systems. Price signal generation using the FF is discussed in the next section.

\begin{figure*}
\centering
\includegraphics[width=0.9\textwidth]{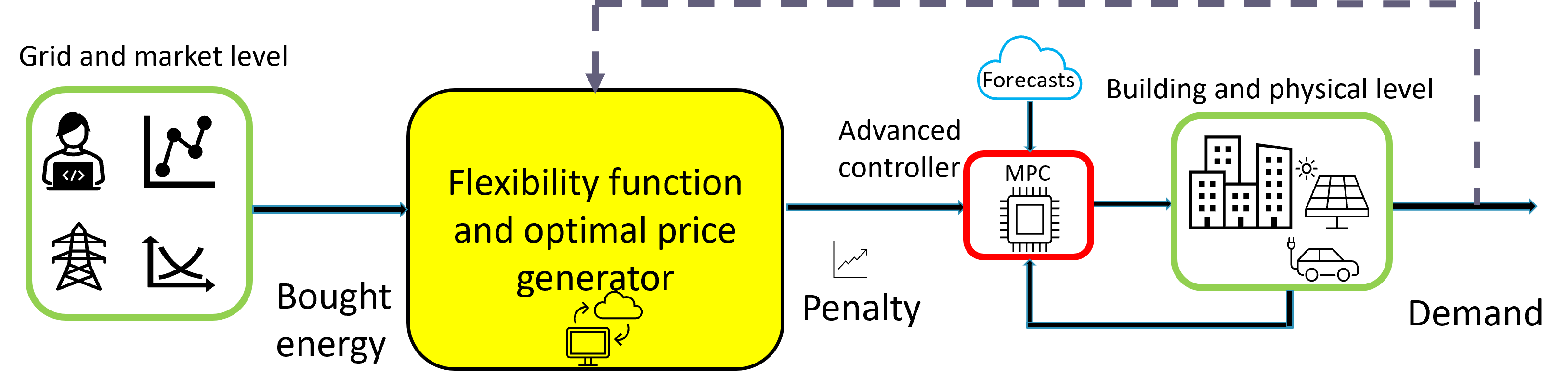}
\caption{Block diagram of higher and lower levels of smart energy OS in the presence of FF.}
\label{Fig:BlockFF}
\end{figure*}
\section{Optimal price generation using FF}\label{sec:OptPrice}

This section is dedicated to the methodologies for the employment of the FF in optimization algorithms to generate the optimal price (penalty) signals.  
The nonlinear FF, as introduced by \cite{Rune20}, is a stochastic dynamical system and is introduced in (\ref{eq:1}).  The dynamics of FF utilize the electricity price and the baseline demand as inputs and provide a prediction for the energy demand. Therefore, it can be considered as a function of the form
\begin{align}\label{eq:2}
    Y_t = FF(U_t, B_t).
\end{align}

The proper form of FF (\ref{eq:2}) makes it an appropriate option to be employed in optimization algorithms. For instance, there are many players in modern energy systems like DSOs, TSOs, and 
aggregators, that can profitably make use of it. As shown in Figure \ref{Fig:SEOS},  aggregators are in connection with the DSO, TSO, BRP, the market (higher levels), and the consumers of a specific area or neighborhood (lower levels). Their intermediate role is an opportunity to connect the advanced controller to the higher levels of the Smart Energy OS. Every day, an aggregator purchases a specific amount of energy for each hour based on the forecasts, baseline demands, and the demand predicted by FF. Once the bought energy is determined, a smart scheduling algorithm is required to manage the consumption, so that the demand does not exceed the purchased energy. This can be done using an advanced controller like a model predictive controller that uses a penalty (price) signal to shift the energy consumption. The block diagram consisting of higher and lower levels of Smart Energy OS in the presence of FF is illustrated in Figure~\ref{Fig:BlockFF}.

The optimal price can then be generated by utilizing the identified nonlinear FF (\ref{eq:2}) in an optimization problem:
\begin{align}\label{eq:3}
\underset{U_t}{\text{minimize}}\ \ \mathcal{J}\left( FF(U_t, B_t)-D_{ref_t}\right),
\end{align}
where $D_{ref_{t}}$ is the amount of energy bought by the aggregator, i.e., it can be considered as a reference to be followed, and $\mathcal{J}$ is a cost function. The optimization problem should be solved for each individual hour, $t = \{1, ..., N\}$, where $N$ is the number of upcoming hours for which energy is bought. The optimization problem (\ref{eq:3}) finds the optimal price ($U_t$) for each~$t$ that minimizes the difference between the bought energy ($D_{ref_{t}}$) and the predicted demand ($FF(U_t, B_t)$) for each~$t$. Note that, at any given time, the predicted demand, $Y_t$, depends indirectly on previous price signals, reference demands, and baseline demand through the flexibility state, $X_t$. The price signal ($[U_1, ..., U_N]$) can then be employed in a model predictive controller formulation for load shifting and demand management.
Suppose that the bought energy for each hour is provided every 24 hours by an aggregator. Then, (\ref{eq:3}) can be used to generate the optimal price/penalty signal for periods of 24 hours. For the first hour, by having $B_1$ and $D_{ref_1}$ in hand, (\ref{eq:3}) calculates $U_1$. The procedure continues until $U_{24}$ is calculated. Then the price signal $[U_1, U_2, ..., U_{24}]$ is transferred to the controllers of the underlying flexible assets which can then be used in their computations.

Solving the optimization problem (\ref{eq:3}) determines the best price, $U_t$, according to the given $B_t$, $D_{ref_{t}}$ and the cost function $\mathcal{J}$ for each $t$. This approach for generating optimal price signals is called \textit{sequential price generation}. This approach has low computational complexity and is an appropriate approach when the computation time must be prioritized, e.g., when generating the price signal for a large number of flexible assets with individual flexibility functions.

\begin{figure*}
\centering
\includegraphics[width=0.9\textwidth]{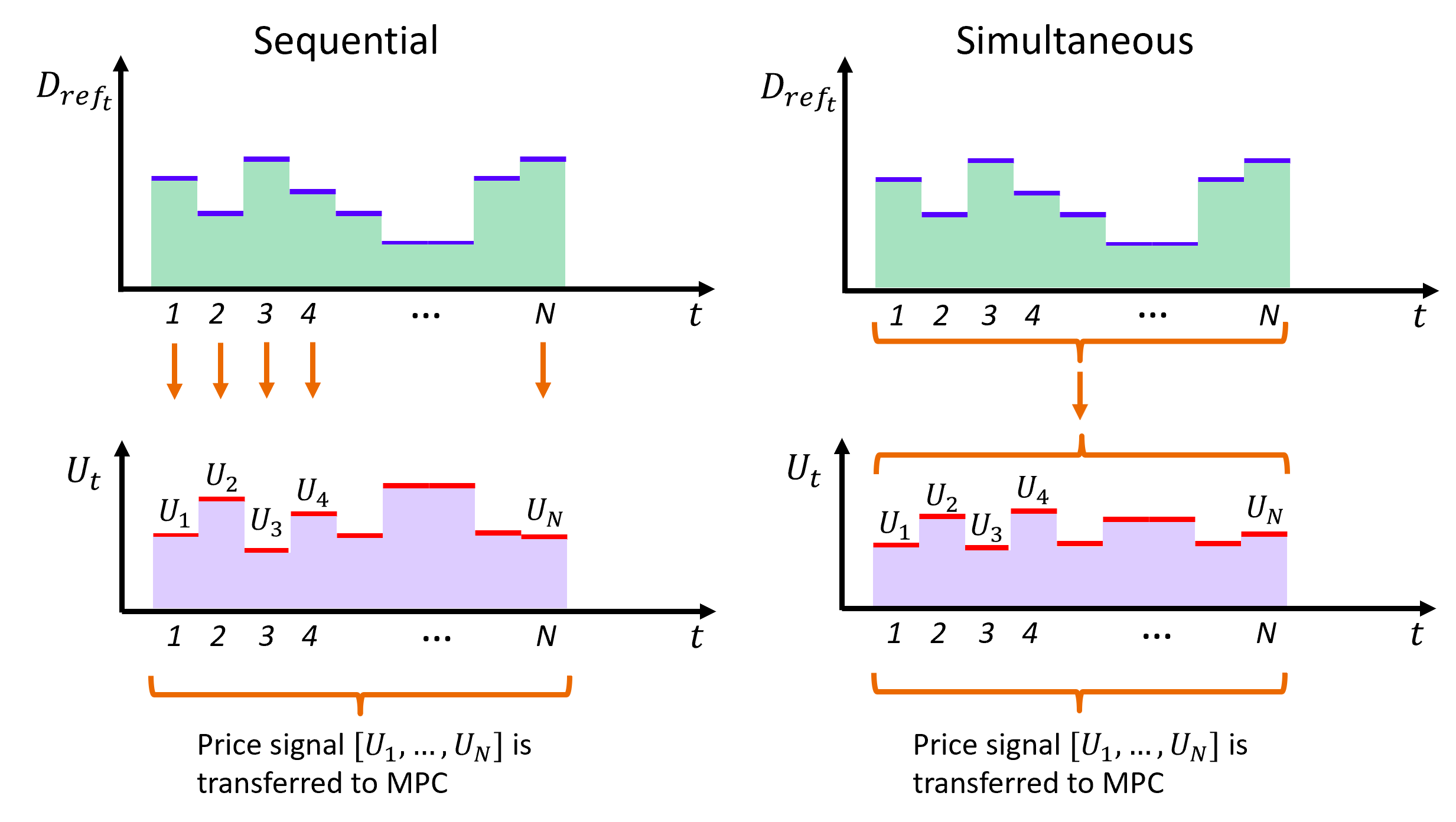}
\caption{Sketch of sequential and simultaneous price signal generation in the Smart Energy OS based on FFs. For given price signals, $U_t$ or $[U_1, ..., U_N]$, the flexibility function is used to predict the demand, $D_t$ or $[D_1, ..., D_N]$, which should be as close as possible to the reference demand, $D_{ref_t}$ or $[D_{ref_1}, ..., D_{ref_N}]$.}
\label{Fig:opt}
\end{figure*}

In the sequential approach, the price for the $k$th hour, $U_{k}$, is calculated based on $D_{ref_{k}}$ and $B_{k}$, regardless of future values of $D_{ref_t}$ and $B_t$. Since aggregators provide the information for $N$ hours ahead, all available data can be used in the decision-making. This is done by reformulating the optimization problem as   
\begin{align}\label{eq:4}
\underset{U_1, ..., U_N}{\text{minimize}}\ \ \sum_{t=1}^{N} \mathcal{J}\left( FF(U_t, B_t)-D_{ref_t}\right),
\end{align}
which utilizes the whole $N$-hour ahead information on reference and baseline demand and calculates the $N$-hour optimal price in one shot. This approach for generating optimal price signals is called \textit{simultaneous price generation}. Different from the sequential price generation, the computational complexity of the simultaneous price generation is high, that is, it takes more time to calculate the optimal price signal using the simultaneous approach. However, it leads to a better demand side management with a lower error between $FF(U_t, B_t)$ and $D_{ref_{t}}$, $t = \{1, ..., N\}$. This is due to the fact that the simultaneous approach utilizes more information ($N$ data points based decision making) while the sequential approach utilizes 1 data point at a time for the decision making and repeats this procedure for each hour ($N$ times). Figure \ref{Fig:opt} demonstrates the two optimization approaches.

Choosing an appropriate cost function ($\mathcal{J}$) is important since this influences the performance of the model predictive controller, and consequently, the demand management. It is the designer's choice to use a proper cost function. Two of the most common cost functions are the absolute value and quadratic functions \cite{Thilker12}. The effect of choosing different cost functions is demonstrated in Section \ref{sec:Simul}.

In addition, secondary objectives and constraints can be added to either (\ref{eq:3}) or (\ref{eq:4}) so that the controller can consider many different problems from upper and lower levels of Smart Energy OS, as well as energy management. These problems are discussed in the following sections.

\section{Benefits of using FF on grid balancing}\label{sec:HighLevel}

Grid balancing can be beneficial from the employment of FF in various ways. Among them are the aggregation of price-demand information and ancillary services~\cite{DeZotti18, DeZotti19}.
These concepts are described in the following subsections.

\subsection{FF for aggregating price-demand information}
As given in (\ref{eq:2}), FF, as a mapping between price and the baseline demand to the predicted demand, is capable of providing aggregated information about the price-responsive energy system without any requirement for extra communication channels in the Smart Energy OS setup. This prominent information is required in different levels of the Smart Energy OS from a neighborhood to a district, a city, or even a wider area. Consequently, the setup offers a possibility for coherent spatial hierarchies, such that the flexibility seen by e.g. at the TSO is coherent with the flexibility seen by the local DSO.  By utilization of FF's demand prediction, DSOs, TSOs, and aggregators would benefit from dynamics of demand variation at different hours of the day due to price changes. Consequently, this leads to more efficient energy management during future horizons in energy systems.

\begin{figure*}
\centering
\includegraphics[width=0.9\textwidth]{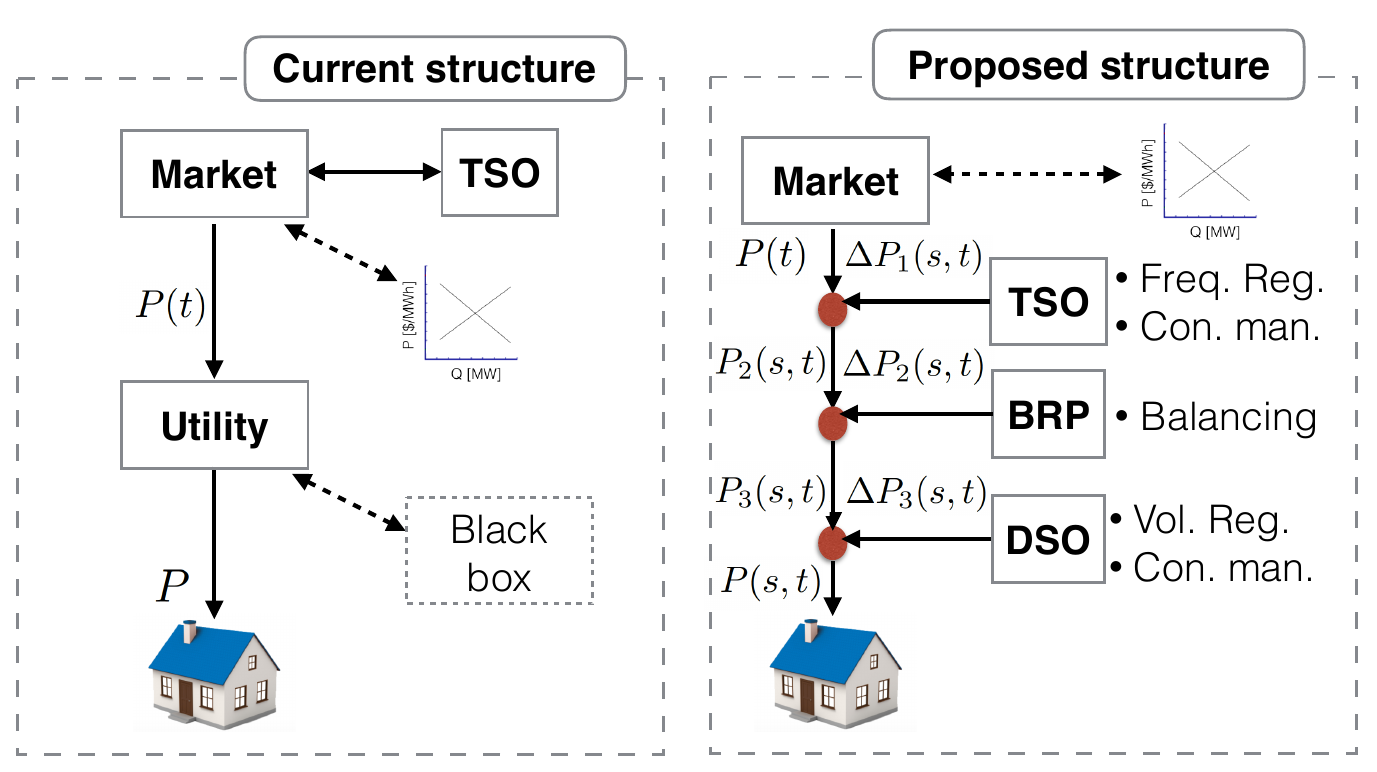}
\caption{The existing vs. novel structures for the price generation considering ancillary services. The supply and demand curve is shown as a tool for electricity market price determination, where $P$ is the price per MWh and $Q$ is the supply power.}
\label{Fig:PriceGen}
\end{figure*}

\subsection{FF for ancillary services}
Another important application of FF is to consider ancillary services for the grid \cite{Rune18}. Ancillary services are the services for guaranteeing equilibrium between supply and demand in an electricity grid \cite{DeZotti18, DeZotti19}. The imbalance in the electricity grid occurs in voltage, frequency, etc., and should be compensated immediately. Employment of FF in the Smart Energy OS equips the energy system with a tool for taking care of these types of imbalances continuously.

This is possible by identifying a function mapping the demand to voltage, frequency, etc. to be used in an optimization problem, and then, finding an optimal price/penalty signal which is capable of triggering a penalty-based controller. To this end, one may define the optimization problem for voltage regulation as
\begin{align}\label{eq:5}
\hspace{-0.5cm}\underset{U_t}{\text{minimize}}\ \ \mathcal{J}_v\left( \mathcal{H}(Y_t)-v_{ref_t}\right)+ \mathcal{J}_U\left( U_t-U_{ref_t}\right),
\end{align}
where $Y_t$ is the predicted demand, defined in (\ref{eq:2}), $\mathcal{H}$ is a function mapping the demand to voltage, $v_{ref_t}$ is the reference voltage, and $U_{ref_t}$ is a nominal price. The first term, $\mathcal{J}_v$, describes the cost of violating the reference voltage, and the second term, $\mathcal{J}_U$, introduces the cost for deviating the price from its reference value. A similar optimization problem can be introduced for frequency regulation.

The ancillary services using FF can be provided along with the optimal price generation. This is done by integrating (\ref{eq:3}) and (\ref{eq:5}) as
\begin{align}\label{eq:5x}
\underset{U_t}{\text{minimize}}\ \ &\mathcal{J}\left( FF(U_t, B_t)-D_{ref_t}\right) \notag \\
&\hspace{-0.8cm}+\mathcal{J}_v\left( \mathcal{H}(Y_t)-v_{ref_t}\right)+ \mathcal{J}_U\left( U_t-U_{ref_t}\right).
\end{align}

Different from (\ref{eq:3}) where the price signal is generated regardless of the grid balancing requirements, this approach considers the grid balancing as well as minimizing the difference between the bought energy ($D_{ref_{t}}$) and the predicted demand ($FF(U_t, B_t)$). A similar optimization problem can
be introduced for frequency regulation.

Solving the optimization problems (\ref{eq:5}) and (\ref{eq:5x}) finds the best price, $U_t$, sequentially. The simultaneous approach can also be used to generate the optimal price as
\begin{align}\label{eq:6}
\underset{U_1, ..., U_N}{\text{minimize}}\ \ &\sum_{t=1}^{N} \mathcal{J}\left( FF(U_t, B_t)-D_{ref_t}\right) \notag \\
&\hspace{-0.9cm}+\mathcal{J}_v\left( \mathcal{H}(Y_t)-v_{ref_t}\right)+ \mathcal{J}_U\left( U_t-U_{ref_t}\right).
\end{align}

The optimization problem formulations discussed in (\ref{eq:5})-(\ref{eq:6}) require information about the demand and voltage relationship. Another possibility would be to eliminate the demand and instead, find a dynamic relation between the price and e.g. the voltage.

Figure \ref{Fig:PriceGen} demonstrates the existing vs. novel structures for the price generation considering ancillary services. In the existing structure, the electricity market is the main decision maker and price signal generator. However, in the novel structure, each service operator is equipped with an FF, based on the duties and regional constraints, and they contribute to determining the price.

\begin{figure*}
\centering
\includegraphics[width=0.9\textwidth]{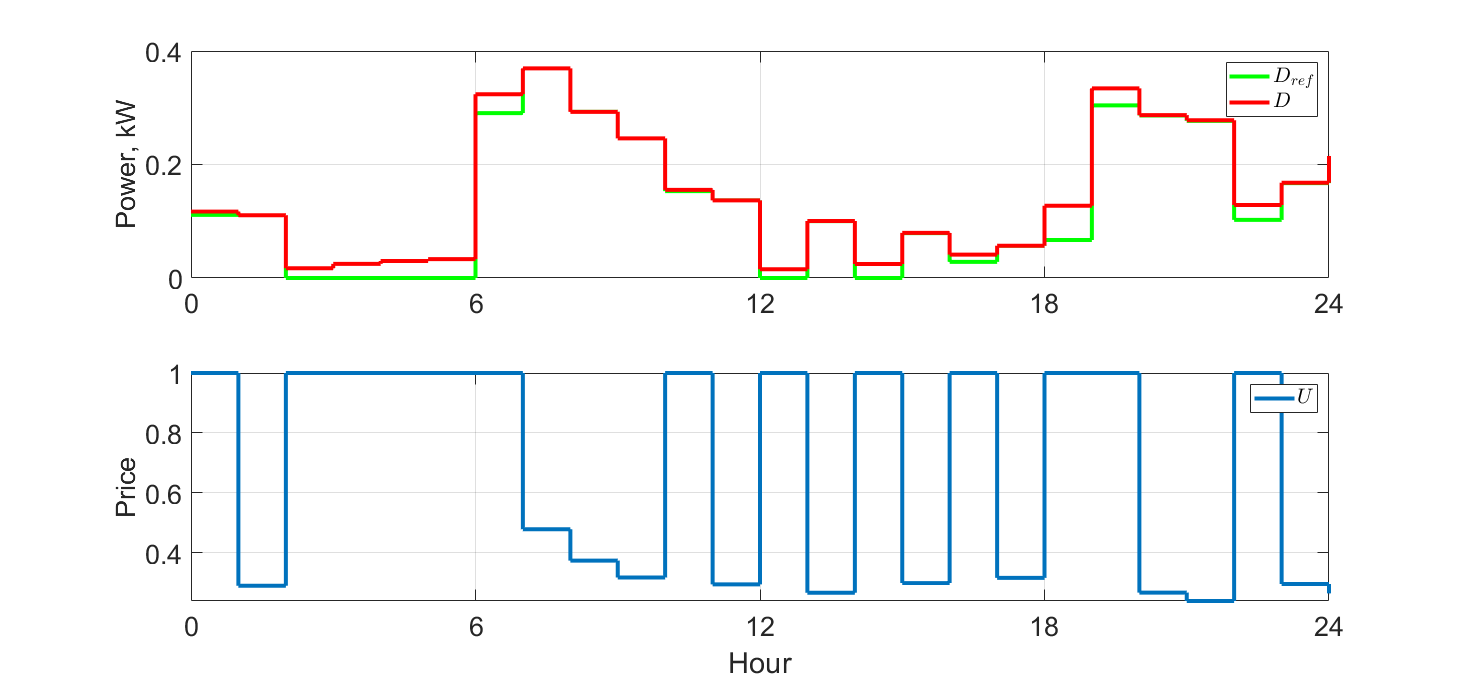}
\caption{Sequential optimal price generation with absolute value objective function. The top panel shows the reference and the actual demand for 24 hours. The bottom panel shows the optimal price signal.}
\label{Fig:seq_v9_abs_one}
\end{figure*}

\begin{figure*}
\centering
\includegraphics[width=0.9\textwidth]{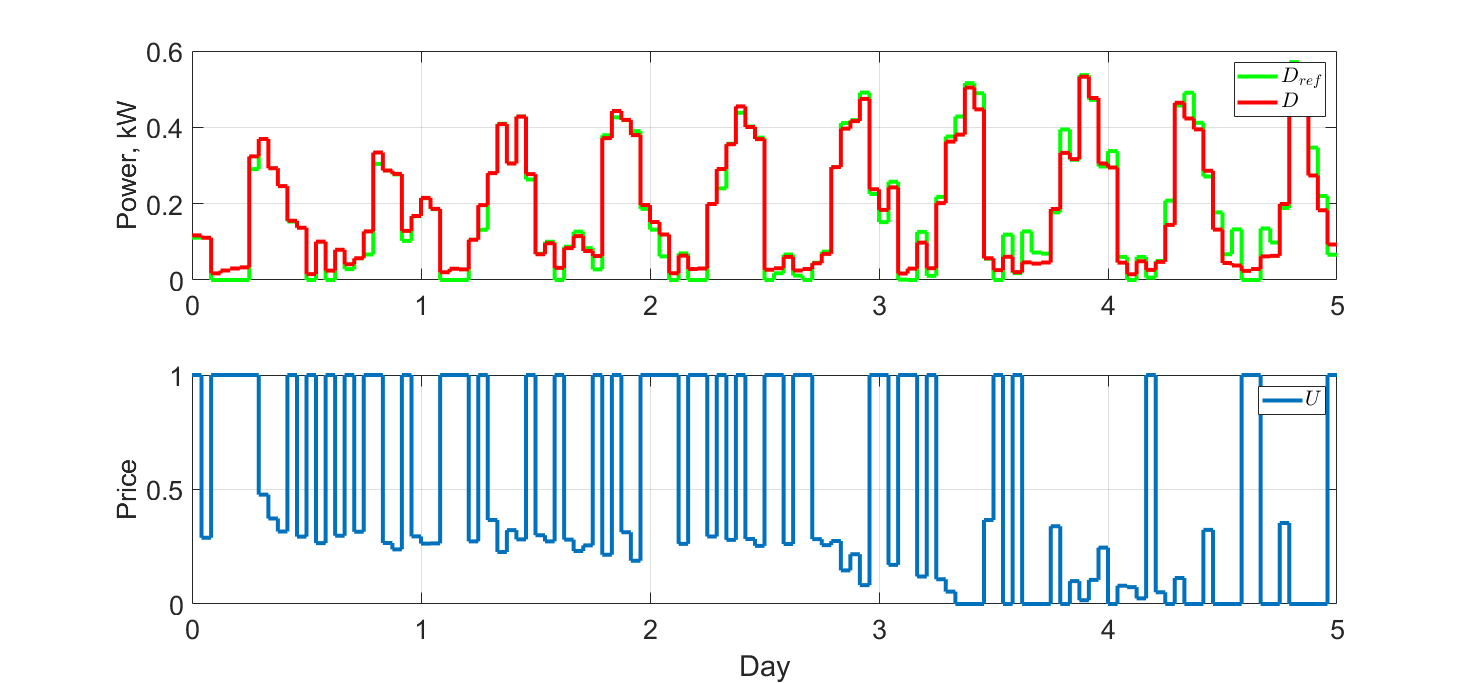}
\caption{Sequential optimal price generation with absolute value objective function. The top panel shows the reference and the actual demand throughout five days. The bottom panel shows the optimal price signal.}
\label{Fig:seq_v9_abs}
\end{figure*}

\begin{figure*}
\centering
\includegraphics[width=0.9\textwidth]{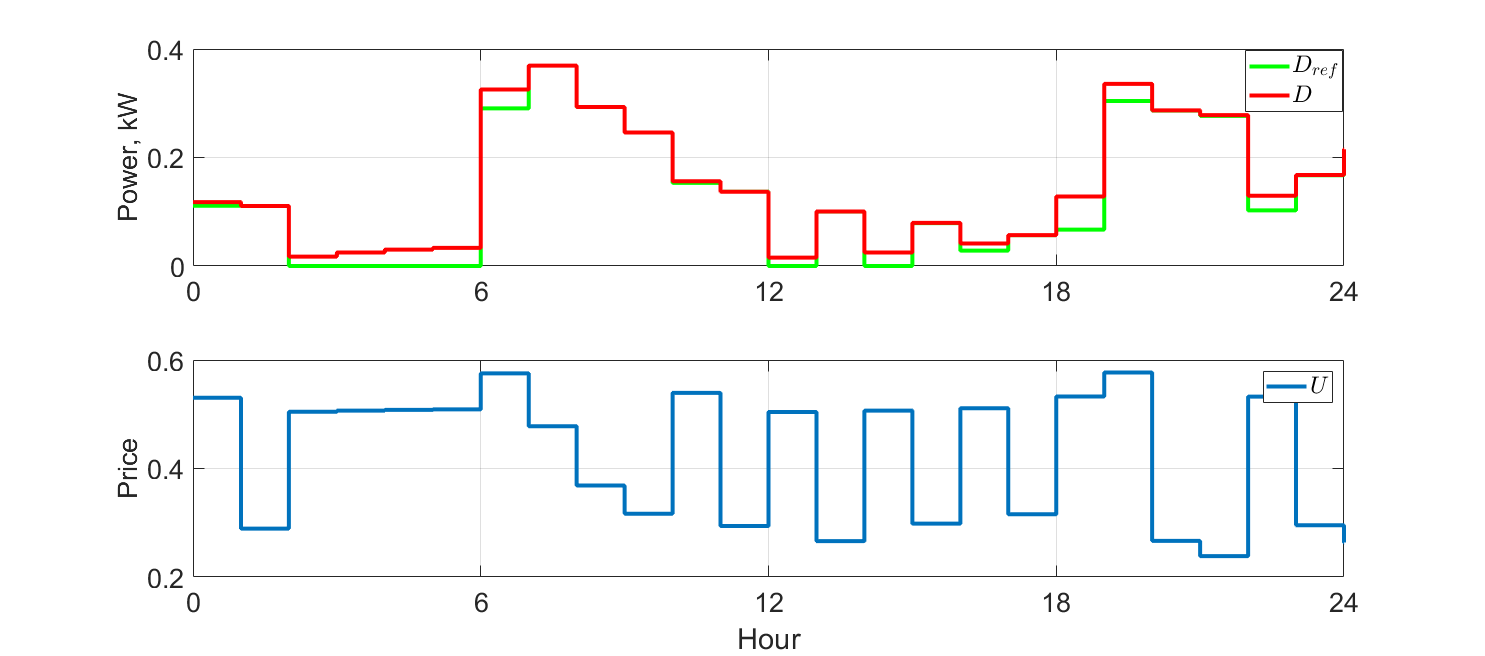}
\caption{Simultaneous optimal price generation with absolute value objective function. The top panel shows the reference and the actual demand for 24 hours. The bottom panel shows the optimal price signal.}
\label{Fig:sim_v10_abs_one}
\end{figure*}

\begin{figure*}
\centering
\includegraphics[width=0.9\textwidth]{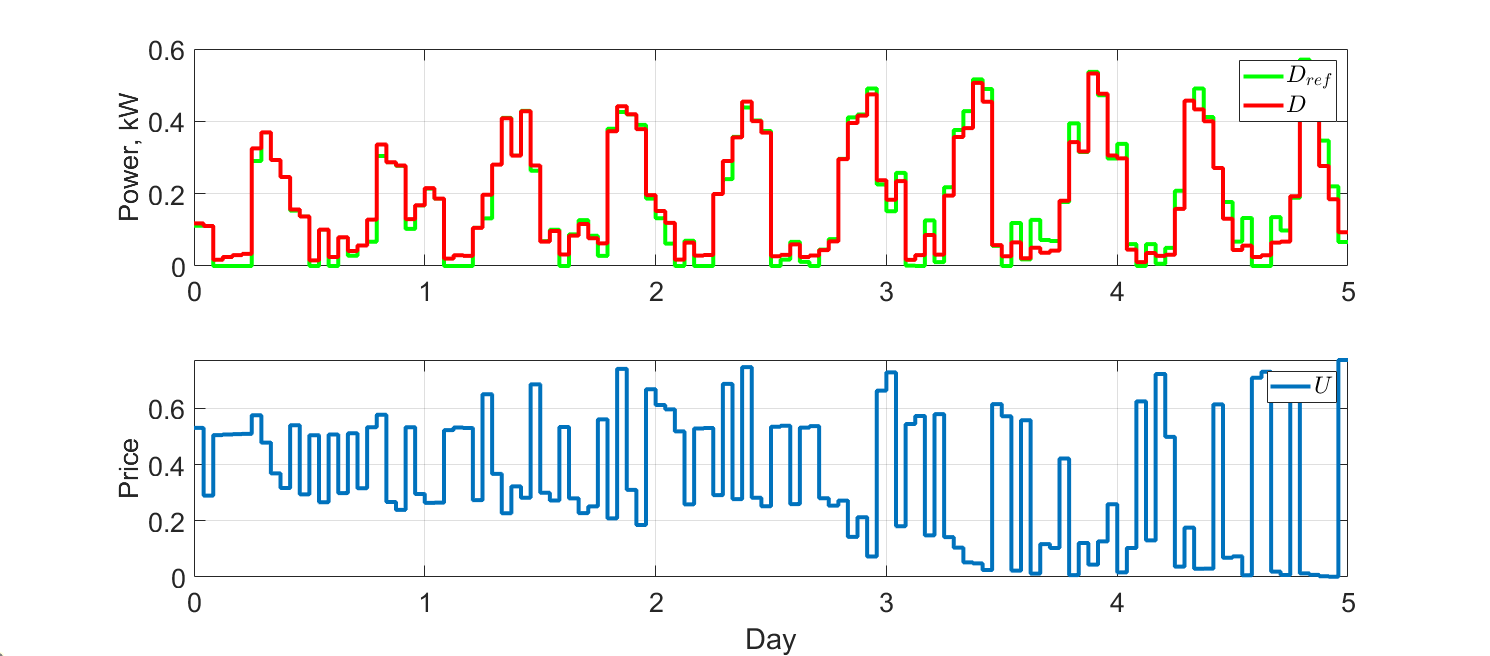}
\caption{Simultaneous optimal price generation with absolute value objective function. The top panel shows the reference and the actual demand throughout five days. The bottom panel shows the optimal price signal.}
\label{Fig:sim_v10_abs}
\end{figure*}

\begin{figure*}
\centering
\includegraphics[width=0.9\textwidth]{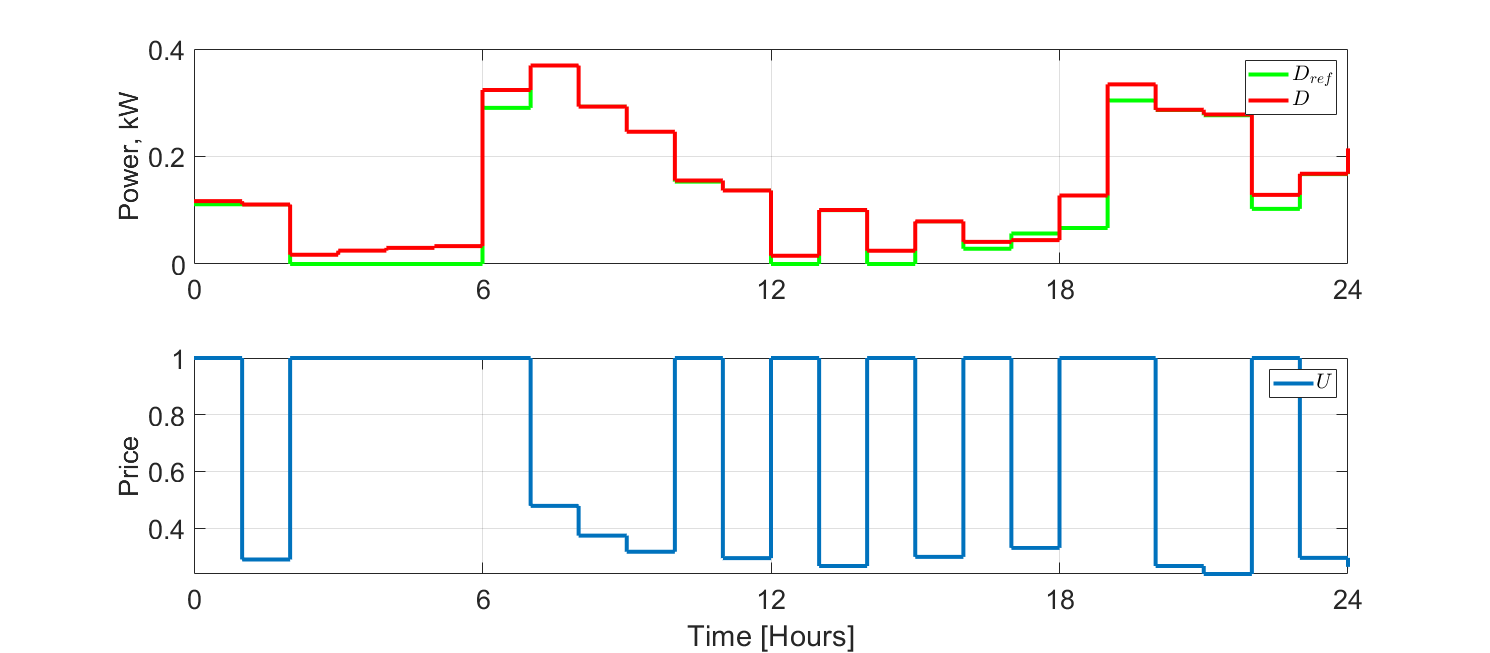}
\caption{Sequential optimal price generation with quadratic objective function. The top panel shows the reference and the actual demand for 24 hours. The bottom panel shows the optimal price signal.}
\label{Fig:seq_v9_quad_one}
\end{figure*}

\begin{figure*}
\centering
\includegraphics[width=0.9\textwidth]{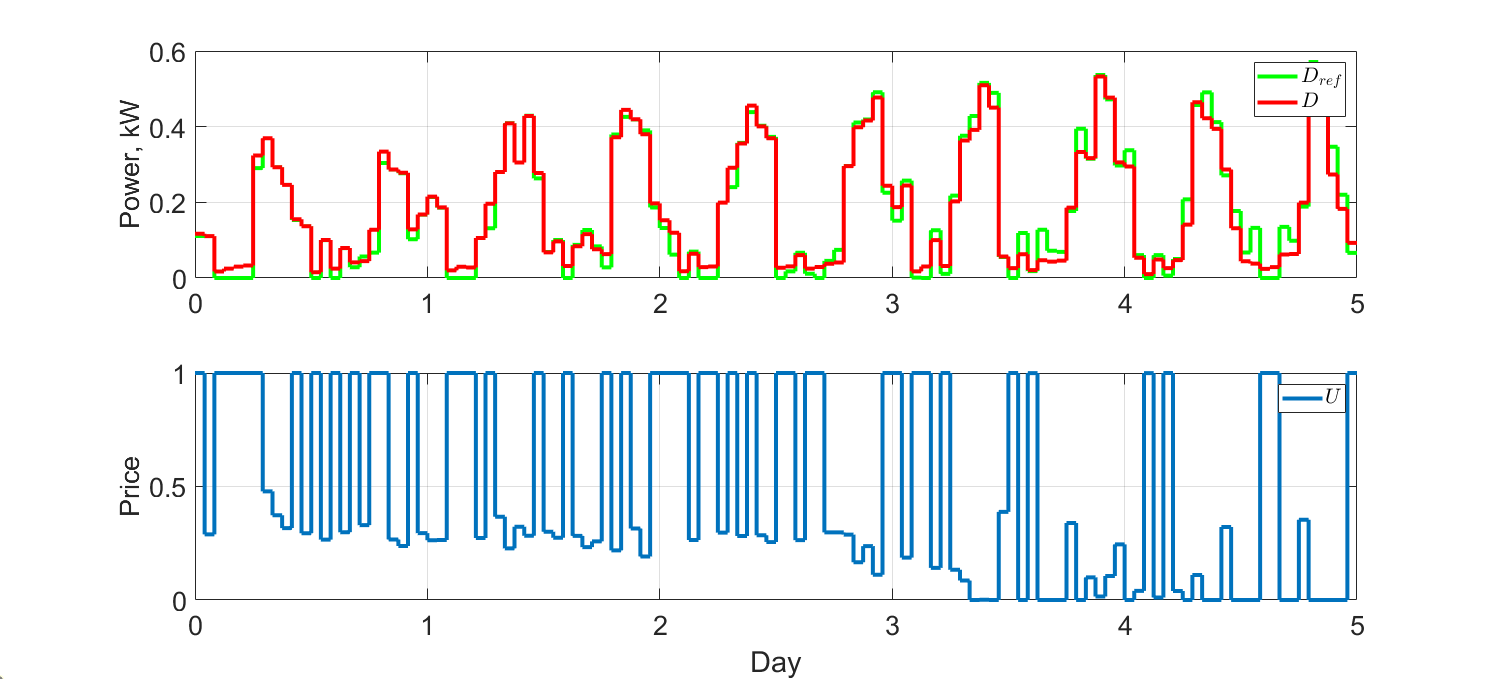}
\caption{Sequential optimal price generation with quadratic objective function. The top panel shows the reference and the actual demand throughout five days. The bottom panel shows the optimal price signal.}
\label{Fig:seq_v9_quad}
\end{figure*}

\begin{figure*}
\centering
\includegraphics[width=0.9\textwidth]{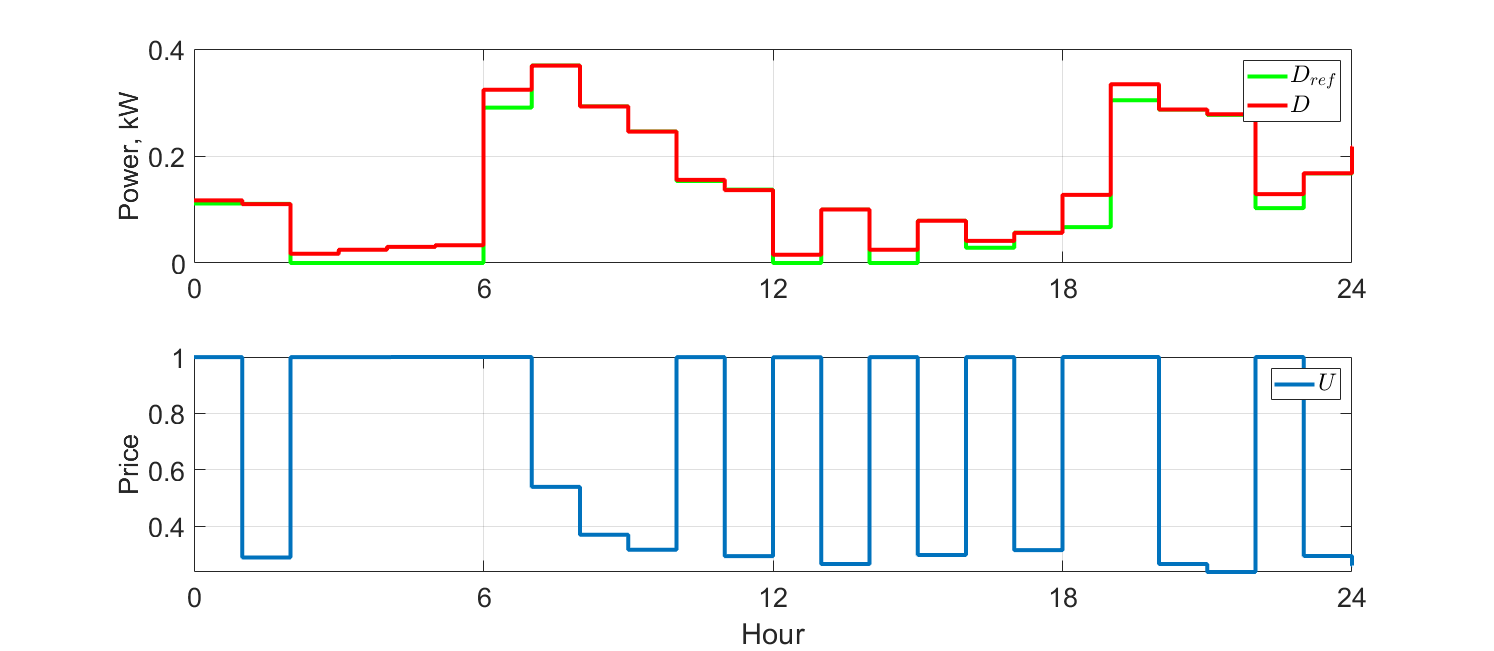}
\caption{Simultaneous optimal price generation with quadratic objective function. The top panel shows the reference and the actual demand for 24 hours. The bottom panel shows the optimal price signal.}
\label{Fig:sim_v10_quad_one}
\end{figure*}

\begin{figure*}
\centering
\includegraphics[width=0.9\textwidth]{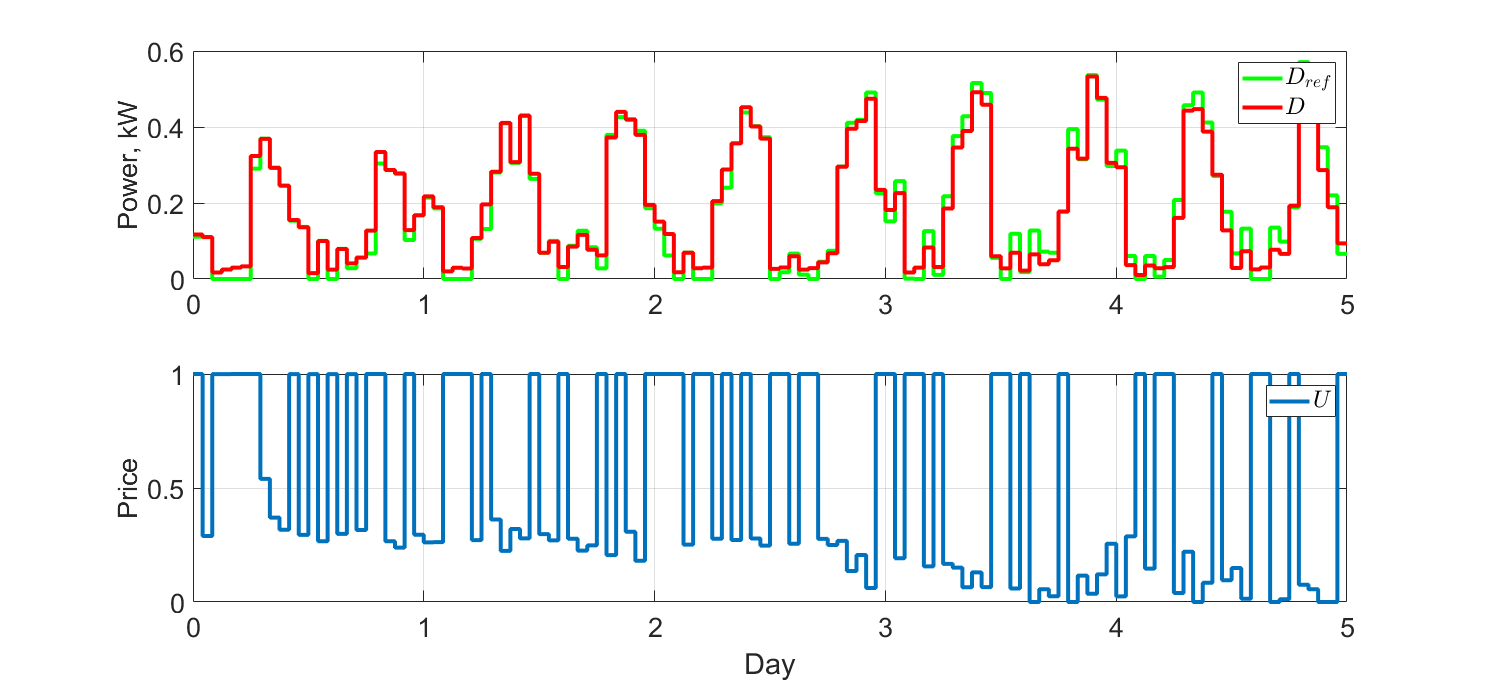}
\caption{Simultaneous optimal price generation with quadratic objective function. The top panel shows the reference and the actual demand throughout five days. The bottom panel shows the optimal price signal.}
\label{Fig:sim_v10_quad}
\end{figure*}

\section{Benefits of using FF on the physical level performance} \label{sec:LowLevel}

Using load shift at the low level of the Smart Energy OS, by e.g., heat pump control has been studied for over a decade, both through simulation \cite{MolitorIEEE2012} and hardware in the loop experiments \cite{MolitorIEEE2013}. While the main application of the FF is demand prediction and energy management, it is still possible to keep the performance and the thermal comfort at the "low level" of the SE-OS \cite{Pal21}. This is done by employing the optimal price generation approach along with an advanced controller like MPC, which is capable of handling state and input constraints.

Another advantage of adopting the FF is related to the cost savings that can be achieved. When using electricity for heating, e.g. through the heat pump, the heat pump buffer storage can be efficiently used for short load shifts. The monetary advantages are related to the variable price of electricity. The end price of the energy is composed of the cost of production, the cost of transportation, and the taxes and fees. As an example, in the Greater Copenhagen Area, the end electricity price in November-December 2023 varied between 0,15 Euros and 0,66 Euros per kWh (source: watts.dk). In the same area, the cost of transportation in the winter season (October - March) is distributed as follows (source: https://radiuselnet.dk):
\begin{itemize}
    \item Low load tariff: 0,021 Euros per kWh between 0:00 and 6:00,
    \item Peak load tariff: 0,18 Euros per kWh between 17:00 and 21:00,
    \item High load tariff: 0,061 Euros per kWh at other times.
\end{itemize}
Time-varying electricity prices along with FF and an advanced controller make it possible to sensibly reduce heating costs \cite{MolitorIEEE2011, Tohidi22D4.2}.

\section{Simulation results}\label{sec:Simul}

In this section, we first generate optimal price signals using different cost functions, i.e. absolute value and quadratic, and different computational approaches, i.e. sequential and simultaneous, and then, use a case study, where a generated price signal is employed for the energy management system.

\subsection{Optimal price generation results}
The bottom panel of Figure \ref{Fig:seq_v9_abs_one} demonstrates the optimal price signal generated using the sequential approach with absolute value objective function, such that the difference between the reference and the predicted demand diminishes (see the top panel). It is seen that the penalty is higher when they are apart from each other. The results for one week are provided in Figure \ref{Fig:seq_v9_abs} to show the price variations in a longer period. The 
simultaneous approach is utilized with absolute value cost function in Figure \ref{Fig:sim_v10_abs_one}. Similar to the sequential approach, the generated price is high when $D$ and $D_{ref}$ have different values. Different from the sequential approach, where the penalty fluctuations are high, the range of change of penalty is smaller. Figure \ref{Fig:sim_v10_abs} demonstrates the results for a longer period. 

Figure \ref{Fig:seq_v9_quad_one} shows the results of the sequential approach for price generation using the quadratic cost function. The top panel illustrates the reference and the predicted demand for 24 hours. The bottom panel provides the generated optimal price signal. The results for five days are also observed in Figure \ref{Fig:seq_v9_quad}. Finally, the results of employing the simultaneous approach with the quadratic cost function for one day and five days are given in Figure \ref{Fig:sim_v10_quad_one} and Figure \ref{Fig:sim_v10_quad}, respectively.

By comparing Figures \ref{Fig:seq_v9_abs_one}-\ref{Fig:sim_v10_quad}, it is seen that different cost functions and different optimization approaches lead to different price signals. For example, depending on the application, it may be desired to employ a price signal with fewer fluctuations. Then, the simultaneous approach with absolute value should be selected. The sum of squares of error between $D$ and $D_{ref}$ is another factor for selecting the price generation method. Table \ref{tab:tab01} compares this factor for different cost functions and optimization approaches. It is seen that the simultaneous approach, which considered $N$ data points, dramatically reduces the sum of squares of error values. Another factor is the penalty signal values. Table \ref{tab:tab02} compares the penalty signals for different cost functions and optimization approaches. It can be observed that the aggregations of penalty signal values generated by the simultaneous approach are much less than the ones generated by the sequential approach. 

\begin{table}
    \caption{Comparison of sum of squares of error between $D$ and $D_{ref}$ regarding different cost functions and optimization approaches.}
    \label{tab:tab01}
    \centering
    \begin{tabular}{| c | c | c | }
    \hline
     & absolute value  & quadratic \\
     & cost function & cost function \\
    \hline
     sequential approach & 2.9 & 3.2\\
     \hline
     simultaneous approach & 0.32 & 0.19\\
     \hline
    \end{tabular}
    \vspace{-5 pt}   
\end{table}

\begin{table}
    \caption{Comparison of sum of penalty signals $U$ regarding different cost functions and optimization approaches.}
    \label{tab:tab02}
    \centering
    \begin{tabular}{| c | c | c | }
    \hline
     & absolute value  & quadratic \\
     & cost function & cost function \\
    \hline
     sequential approach & 1140 & 1160\\
     \hline
     simultaneous approach & 417.8 & 630.6\\
     \hline
    \end{tabular}
    \vspace{-5 pt}   
\end{table}

\subsection{HVAC system control for the demand-side management}

\begin{figure*}
\centering
\includegraphics[width=1\textwidth]{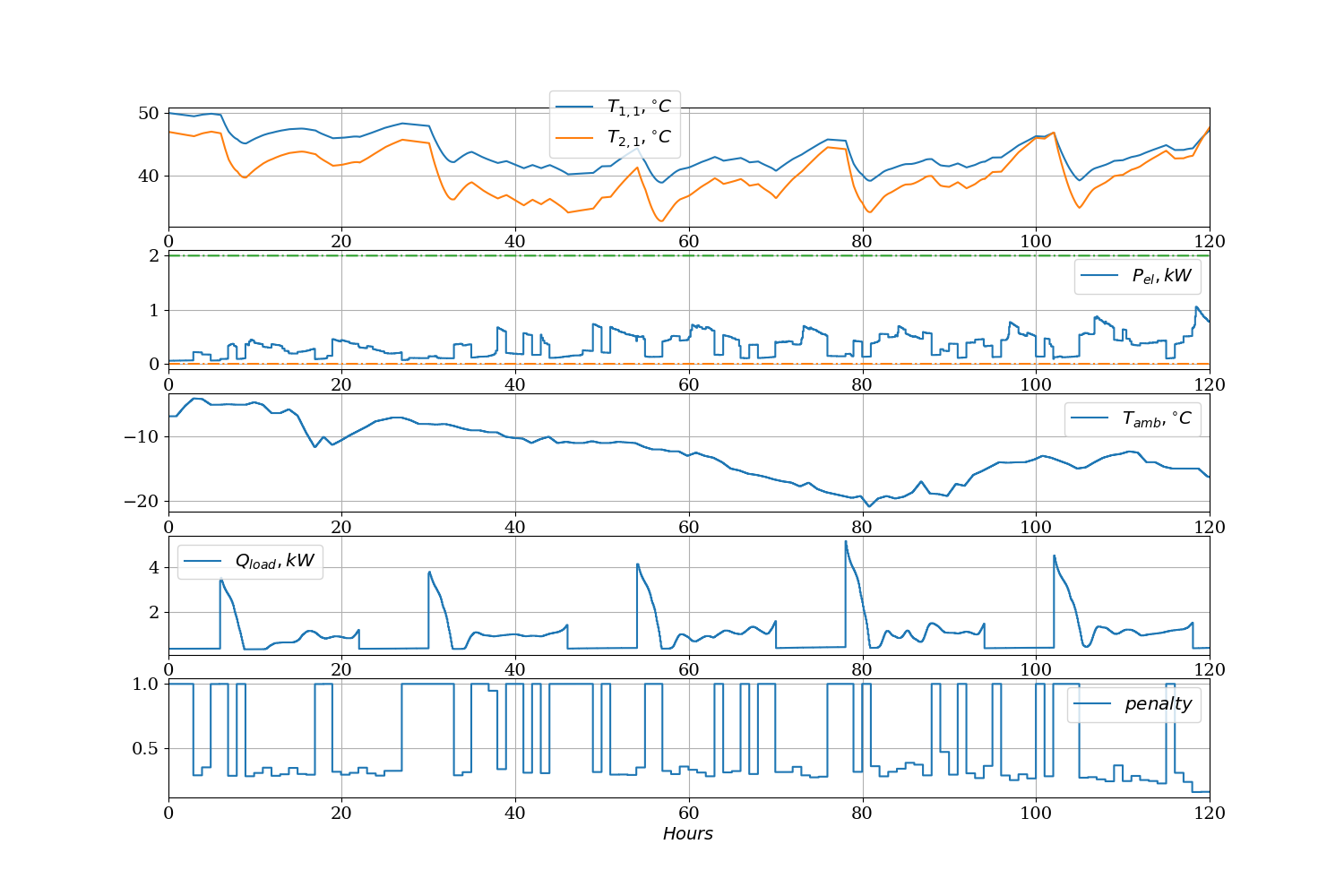}
\caption{Simulation results of the MPC using optimal penalty signal. The top panel shows the water temperature of
the top and bottom of the tank. The second panel demonstrates the electricity consumption. The
ambient temperature is provided in the third panel. The requested load of the tank and the optimal penalty signal are shown in
the fourth and fifth panels, respectively.}
\label{Fig:HVACcontrol}
\end{figure*}

\begin{figure}
\centering
\includegraphics[width=0.3\textwidth]{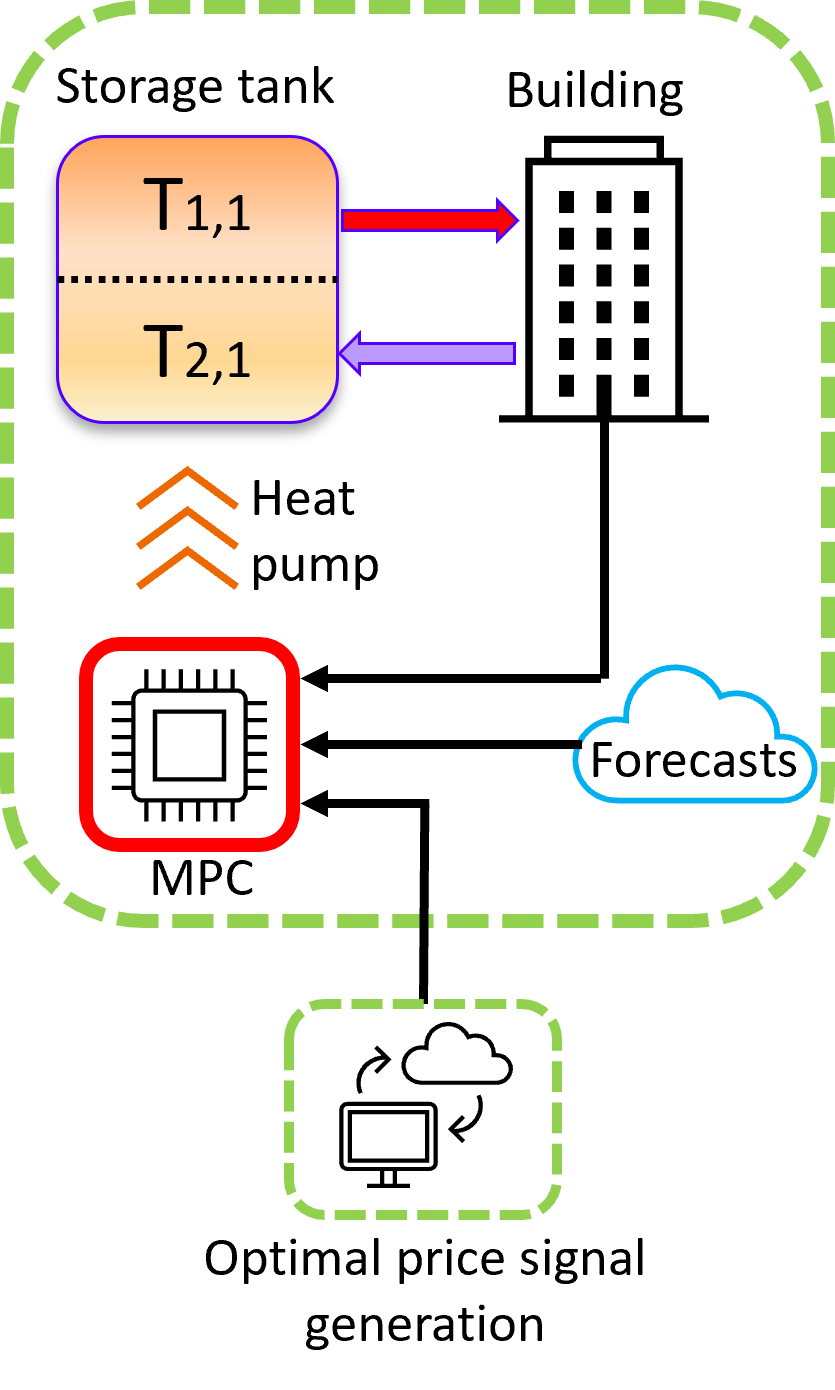}
\caption{Building energy management system with HVAC system control.}
\label{Fig:HVAC}
\end{figure}

As a case study to demonstrate the effectiveness of the optimal price signal for demand-side management, we use the data of a new development in Fredrikstad, Norway. It is the largest development of plus energy houses in Norway and has a strong focus on energy sharing and flexibility in the neighborhood \cite{Tohidi22D4.2, Tohidi23D4.3}. Due to the high flexibility potential of the HVAC system, we aim to control the HVAC system via an advanced controller.

The HVAC system in this neighborhood is simplified and provided in Figure \ref{Fig:HVAC}. As can be seen in Figure \ref{Fig:HVAC}, the HVAC system consists of a heat pump and a storage tank. The heat pump generates the required thermal demand, and the storage tank stores an enormous amount of hot water for the building. A model predictive controller (MPC) has been developed to control the storage tank temperature while taking the demand-side management into account. The controller is equipped with a dynamic model of thermal dynamics to predict the future evolutions of the thermal system. To this end, a grey-box model is identified from a bank of data generated by a white-box model representing the thermal dynamics of the neighborhood \cite{Tohidi22}. In addition, MPC requires a price/penalty signal to handle the demand-side management. In this study, an optimal price signal is generated using the simultaneous approach with a quadratic cost function. 

Simulation results demonstrating the efficacy of the employed controller with optimal penalty signal generation are provided in Figure \ref{Fig:HVACcontrol}. The top panel shows the water temperature of the top and bottom of the tank. The second panel shows the
electricity consumption to heat the water temperature in the tank. The ambient temperature is provided in
the third panel. The requested load of the tank and the optimal penalty signal are shown in the fourth and fifth
panels, respectively. It is observed that the controller shifts the demand to the periods when the penalty signal is lower.

\section{Outlook}\label{sec:Out}
In this paper, we have focused on the use of FFs for computing \emph{electricity} prices. However, it is equally applicable to other energy demands and in this section, we will describe how it can be used to integrate power grids with district heating grids.
In a district heating grid, a central heat plant or combined heat and power plant heats up water which is distributed to residential, commercial, and industrial consumers through a grid of insulated pipes~\cite{Lund:etal:2018}. The heat can be generated using biomass combustion, waste incineration, industrial waste heat (e.g., from data centers or the process industry), renewable energy sources (wind, solar, geothermal, etc.), fossil fuels (such as gas, oil, and coal), or nuclear power.

District heating grids themselves constitute flexible assets that can be used for load shifting and ancillary services in power grids~\cite{Boldrini:etal:2022}. Surplus power production can be used by booster heat pumps to generate heat that can be stored using heat accumulators (for hours or days) or pit thermal energy storage (PTES) solutions (for months or entire seasons). If heat is generated using combined heat and power plants (CHPs), which are common in Denmark~\cite{palsson1993a}, the district heating grid will be even more flexible due to surplus heat generation when the electricity price, and therefore also power generation, is high. Furthermore, as district cooling generates significant amounts of heat~\cite{Ostergaard:etal:2022}, there is also significant potential in combining them with district heating grids. Such combined grids are referred to as fifth generation district heating and cooling grids~\cite{Lund:etal:2021}. As the FF quantifies demand-price relationships, it can also quantify the flexibility of large-scale flexible assets such as district heating (and cooling) grids.

Additionally, district heating consumers can be flexible with respect to both their power and heat demand. However, in conventional markets, it is not possible to offer this flexibility in both markets simultaneously. In contrast, in the Smart Energy OS, a FF can be identified for both the power and heat demand and the optimization-based approach described in this paper can be used to generate separate electricity and heat price signals in order to indirectly control both demands. As both the power and heat consumption are envisioned to be managed automatically by a smart energy management system (e.g., based on economic MPC), it is straightforward to account for both prices at the same time. By exploiting this flexibility and predicting the heat demand, it is possible to reduce the supply temperature and thereby also reduce the heat losses in the district heating grid. See~\cite{Lund:etal:2022} for a discussion of motivation tariffs for district heating consumers and how they should be implemented to lower the supply temperature. Finally, for lower supply temperatures, more types of heat sources can be used, which enables a higher degree of sector coupling, i.e., a higher level of industrial waste heat utilization.

\section{Summary}\label{sec:Sum}
In this work, the hierarchical structure of the Smart Energy OS along with its main components, from the market level to the physical level, have been introduced. Nonlinear FF and demand predictability have also been described. Furthermore, optimization problems with different formulations, i.e. sequential and simultaneous, are proposed for optimal price signal generation. The benefits of employing FF on grid balancing through aggregating price-demand information and ancillary services have also been introduced. These results have been extended to consider optimal price generation as well as providing ancillary services. Moreover, the benefits of FF on the physical level (low level) of the energy system performance have been discussed. The possibility of deployment of FF for the district heating grids has also been discussed. The simulation results established in this paper demonstrate the efficiency of utilizing FF for demand-side management as well as its capability for linking the market and physical levels of Smart Energy OS. 

\section*{Acknowledgments}
The authors gratefully acknowledge  \textit{Sustainable plus energy neighbourhoods (syn.ikia)} (EU H2020 No. 869918), \textit{ELEXIA} (EU Horizon Europe No. 101075656), \textit{ARV} (EU H2020 No. 101036723), and  \textit{SEM4Cities} (Innovation Fund Denmark, No. 0143-0004).

\printcredits

\bibliographystyle{unsrt}

\bibliography{cas-dc-template}



\end{document}